# RANSlicing: Towards Multi-Tenancy in 5G Radio Access Networks

Wardah Saleh and Shahrin Chowdhury

Department of Computer Science, American International University, Bangladesh

*Abstract*

*A significant purpose of 5G networks is allowing sharing resources among different network tenants such as service providers and Mobile Virtual network Operators. Numerous domains are taken in account regarding resource sharing containing different infrastructure (storage, compute and networking), Radio Access Network (RAN) and Radio Frequency (RF) spectrum. RAN and spectrum, transport. Spectrum sharing and RAN are anticipated as the fundamental part in multi-tenant 5G network. Nevertheless, there is a shortage of evaluation platforms to determine the number of benefits that can be acquired from multilevel spectrum sharing rather than single-level spectrum sharing. The work presented in this paper intend to address this issue by presenting a modified SimuLTE model is used for evaluating active RAN based on multi-tenant 5G networks. The result shows an understanding into the actual advantages of RAN slicing for multi-tenants in 5G networks.*

## 1. Introduction

Last few years have observed a remarkable expansion in data traffic with the introduction of IoT (Internet of Things) devices such as the smartphones and tablets, especially due to content rich multimedia and cloud applications and the upcoming vertical market services in automotive, ehealth etc. [1]. This results the mobile network operators (MNOs) to constantly provide their network structure and infrastructure which in turn rises the service and resource cost. In near future, the usage of global mobile data traffic will be increasing due to high usage of mobile users. Considering such high traffic demand will need severe change to how networks operate today. Moreover, the Fifth Generation (5G) networks will have to expand its efficiency far beyond the previous generation mobile communication. The idea of 5G is to offer connections that are myriad faster than current connections along with average download speeds of around 1 Gbps expected to soon be the benchmark where it is manifested to be deployed by 2020 across the world. The most relevant requirements of 5G networks in which this paper aimed to contribute up to 10 Gbps data rate with a minimum of 50 Mbps download speed and 100 Mbps upload speed even in the worst-case scenarios; then there comes 1 ms latency to provide seamless connectivity to the IoT devices and 99.999% availability [2] [3]. The capabilities of 5G together with other quality of service (QoS) features will be performed with the development of LTE in combination with new radio-access technologies like Software-Defined Networking (SDN), Network Function Virtualization (NFV) etc. The main problem withtraditional network architecture is that it is not dynamic; users do not have much control and it just contains current networking needs of the user. That means, the scope for innovation is too limited with respect to network controlling, virtualization and fault tolerance etc. which is why SDN has emerged to solve the mentioned problems. On the other hand, network slicing can provide support on-demand tailored services for different application scenarios simultaneously using the same physical network by slicing a physical network into various logical networks. With the help of network slicing, networks resources can be dynamically and efficiently assigned to logical networks slices according to the corresponding QoS (Quality of Service) demands. Moreover, SDN controllers can control network slicing in a centralized manner. Virtualization is a key process for network slicing as it enables effective resource sharing among slices. So, the idea of SDN and NFV targets at making an active network infrastructure that makes it easier to bring new services to the end users in a time and resource efficient manner. Moreover, it facilitates the facility providers and MNOs to use shared (multi-tenant) infrastructure. A multi- tenant 5G network looks forward to a slicing architecture that dynamically provides virtual 5G network slices addressing specific tenant requirements and virtualized SDN/NFV control instances for a full tenant control of allocated virtual resources [4]. From the perspective of 5G networks that is multi-tenant based; comprise the RAN and RF spectrum for sharing important resources. These spectrums introduced different models for distributing resources for example, spectrum leasing, mutual renting and co-primary sharing. Also, the work on spectrum sharing in multi-tenant 5G Radio Access Network (RAN) has been done [5]. But in the past, the network was sliced outside the LTE protocol stack [6]; which does not really virtualize the RAN. In [5], the authors have showed a way of RAN virtualization and fined-grained spectrum sharing but according to our research study RAN can further be sliced or virtualized. Besides, only single level slicing and sharing is done in their work. The authors have also done long observation periods to make sharing decisions where spectrum is getting wasted and our research made a significant improvement in their spectrum sharing techniques. So, this paper focuses on multi-level active spectrum sharing on 5G Radio Access Network (RAN) leveraging the wasted radio resources.

## 2. RELATED WORK

Previously there have been several research-works on performing the cognitive radio concepts of active RAN sharing techniques in 5G networks. In [7], the authors show a technique of partial resource reservation of active RAN sharing which is flexible and permits individual operator to definite a least possible share of resources at the same time; first-come-first-served technique is applied for accessing shared common resources. A technique named as partial resource reservation is proposed in the paper which is used for addressing the scheduler as well as admission control and Long-Term Evaluation (LTE) networks. Their system level simulation showed a comparison between full reservation scheme and their anticipated scheme that can compliantly distribute based on their traffic priorities and their actual traffic loads for the shared resources operation. Consequently, develop the spectrum usage and the network owner's revenue both are compared to full sharing scheme. Therefore, this scheme can maintain least definite performance for individual operator.

According to [6], the authors presented a design and implementation of cell slice for slicing wireless resource in a cellular network for effective Radio Access Network (RAN). CellSlice is a gateway-level solution that achieves the slicing without modifying the base stations' MAC schedulers, thereby significantly reducing the barrier for its adaptation. The most challenging job here is to achieve slicing with a gateway-level solution though at fine timescales, resource sharing decisions occur at the base stations and at the gateways these decisions are invisible. There are two directions: uplink direction and downlink direction. By using a simple feedback based adoption algorithm, CellSlice overcomes the challenge by indirectly constraining the uplink scheduler's decision in the uplink direction. On the other hand, for downlink, the authors have implemented a technique used by NVS (Native base station Virtualization Solution) and presented that effective downlink slicing can be easily accomplished without modifying base station schedulers. The authors presented a model of CellSlice on a Picochip WiMAX testbed.

The CellSlice's performance for both remote uplink and downlink slicing has demonstrated through both prototype interpretation and simulations that is close to that of NVS. The architecture of CellSlice is access-technology independent and thus can be symmetrically applicable to LTE, LTE-Advanced and WiMAX networks.

Paper [5] is based on 5G network algorithms of spectrum sharing. The paper presents an approach which is centralized and supports active RAN those are fine-grained. To analyze this approach an open-source system level LTE/LTE-A User-Plane simulation model for OMNET++ which is a modified smiLTE model has been used. For Analyzing the active RAN, the mentioned model can be utilized along with spectrum sharing models which are treated in a 5G network that supports multi-tenant feature. Resource scheduling functions and dedicated spectrum of RAN scheduling that is allocated dynamically is permitted by root modules. The authors illustrated exploratory experimental results that provides, a specific observation focusing on the real advantage and in different time-frequency algorithms RAN tenants allow exchange of active spectrum among themselves. In different time-frequency granularities; dedicated spectrum of a RAN tenant can be allocated to another tenant with permission. They considered exchange of different time-frequency granularities and single cell containing two tenants to investigate the potential benefits. End-to-end solution with more complicated distribution in a dynamic slice creation is their future work.

## 3. NETWORK ARCHITECTURE

A simulation platform is built which is capable of dynamically virtualizing the RAN (Radio Access Network). For this, a multi-tenancy supporting system level simulator has built in a widely used platform (e.g. OMNet++) to implement multi-level RAN slicing. After building the simulation platform, a controller is built on top of the slices to implement spectrum sharing techniques.

### 3.1. RAN Virtualization

An oversimplified definition of Radio Access Networks or RAN would be the base stations we see in our daily life. It comprises of Antennas and radio resources or spectrum. Imagine a scenario, where Mobile Network Operators (MNOs) like GP, Banglalink and Airtel could reside in the same base station even though they are independent of each other. They have their own spectrum, users and set of rules. But there is a centralized controller on top of them that can control them and share each other's physical or radio resources when not being used. This is pretty much how we have virtualized the RAN.

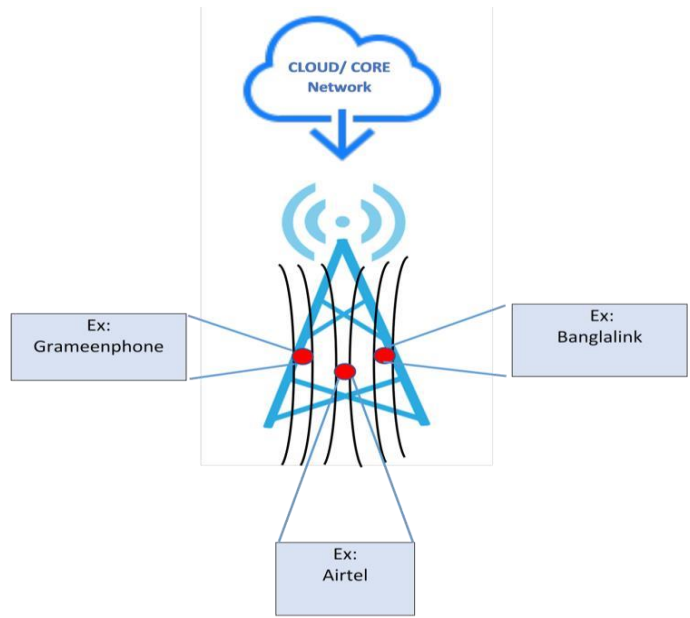

Figure 1. RAN Virtualization

The above figure is a scenario of RAN virtualization where it creates an adaptive and flexible network providing flexibility and scalability for Mobile Network Operators (MNOs). Here, MNOs can co-exist virtually in the same RAN yet, they can act independently and share resources if necessary. While early implementations provided each physical BS (Base Station) with its own dedicated computing resources resulting in an over-provisioning of computing resources, more advanced implementations which is done in this thesis that allows a dynamic reassignment of processing resources to BSs based on the requirement [8]. Hence, a platform is needed that lies at the intersection of real-time architectures for handling communication signals and large-scale information processing systems by the virtualization and centralization of the RAN. This RAN sharing architecture is expected to permit distinct core network operator to join with a shared radio access network. The operators along with sharing radio network elements are possibly expected to share radio resources among them. Therefore, network sharing is an adjustment between operators and the user should have clear regarding this. This designates that a supporting UE (User Equipment) requires the eligibility to distinguish between operators of core network who are existing in a shared radio access network and these operators can be maintained in the similar way as non-shared operators does.

For this, a RAN is virtualized, making it agile and the novelty of the work is to create dynamic slices. To this extent, a simulation platform has built in omnet++ which is a widely used simulation platform for computer networks that supports RAN virtualization on Multi-Tenancy.

## 3.2. Multi-Level Slicing

As per previous discussion, multiple operators like GP or Banglalink can reside inside the same base station. Previously, only single level slicing and sharing is done in [5]. In our platform, the slice of GP can further be divided into more slices. Consider a scenario, GP has only two customers requesting video. GP will allocate all its resources to these customers now. It might be the case that they do not need all of it. In the meantime, another two customers join and requests VOIP calls. Our platform will automatically calculate the required resources for the VOIP calls and allocate it to them. This will ensure a guaranteed performance for the VOIP and video streaming. Once the VOIP call finishes video streaming will enjoy the whole bandwidth again.

So, the focus in this part is to ensure dynamic slice creation and guaranteed bit rate. The following figure is an example of the mentioned scenario.

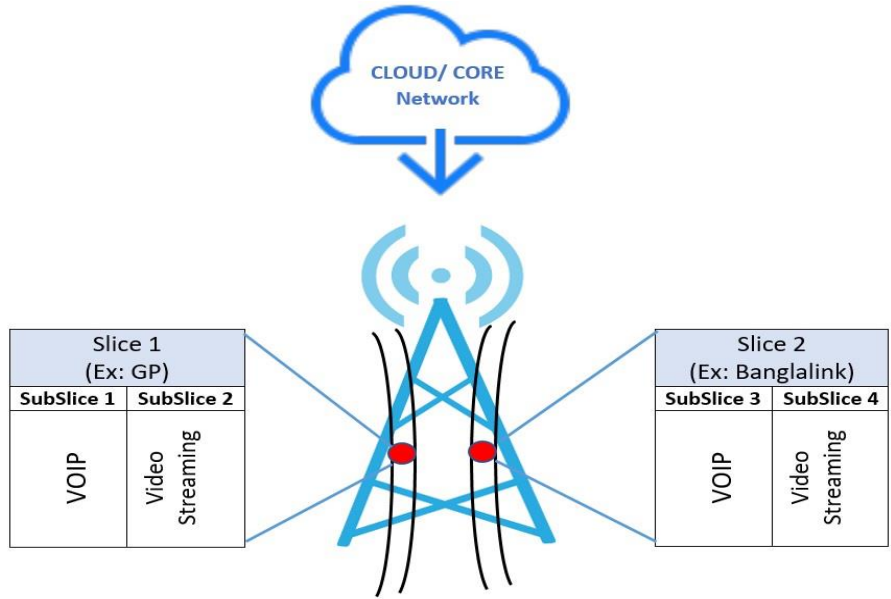

Figure 2. Multi-level slicing in 5G network

## 4. SLICING SCHEME

In this section, we describe the proposed mechanism that is multi-tenancy scheme where the architecture of multi-level slicing has shown. Before describing multi-tenancy scheme, single tenancy is also described that has done previously to understand the basic difference between them.

### 4.1. Single-Level Scheme

In [5], exchange of distinct time-frequency granularities and examine the probable benefits are considered while two tenants residing in a single slice or cell. For this, the concept in SimuLTE has reformed the current node's structure to allow multi-tenant eNodeB that has physical layer resources as well as MAC modules for individual tenant.

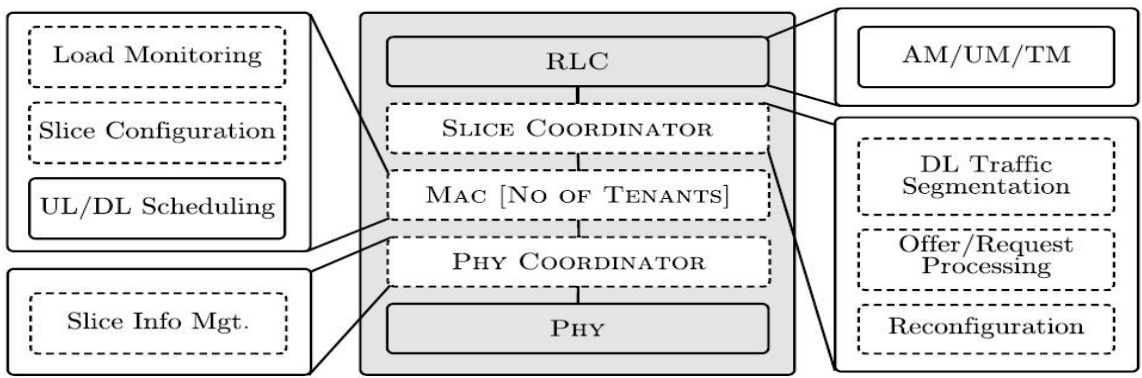

Figure 3. New and improved sub-modules inside eNodeB structure [5]

Figure 3 represents added and reformed sub-modules (drawn with dash line) in LTE-NIC composite nodes. Here, Slice Coordinator and PHY Coordinator are used to coordinate MAC submodules for tenants. These tenants are arranged in a user –defined array which is supported by eNodeB. SimuLTE model modifies the MAC modules for supporting new communication (messages and interfaces) and carry-on discrete set of physical layer resources. The architecture of single-level slicing has shown below where the new modified sub-modules are included.

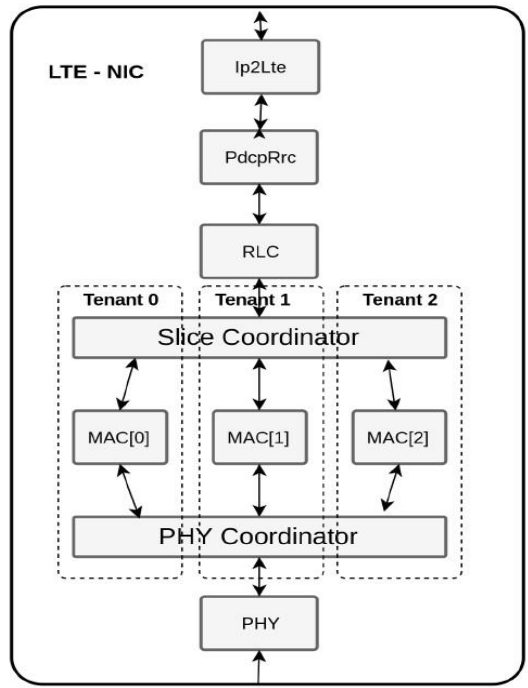

Figure 4. Architecture of single-level slicing [5]

To ensure tenant specific MACs with separate and independent functionalities an array of MACs is used instead of a single MAC in the LTE-NIC module. The array size of the MACs is equal to the number of tenants to be specified by the user. The connection between the MACSlice Coordinator and MAC-PHY Coordinator has to be dynamic with respect to the number of slices. When it receives data packet from upper or lower layer, they are being treated in such a way as if there is only one MAC. They don't communicate within themselves, rather Slice Coordinator and PHY Coordinator controls it.

### 4.2. Multi-Level Scheme

Gartner [9] defines multi-tenancy as "A reference to the mode of operation of software where multiple independent instances of one or multiple applications operate in a shared environment. The instances (tenants) are logically isolated but physically integrated." The Next Generation Mobile Networks (NGMN) alliance anticipates a 5G architecture that influences the structural partition of software and hardware and the programmability offered by Software Defined Networking (SDN) and Network Function Virtualization (NFV). Fundamentally the 5G architecture is an original SDN/NFV architecture covering features ranging from devise (mobile/fixed) framework, all the management function and virtualized functions to organize the 5G system (9). The 5G architecture shall hold a wide range of use cases from the vertical industries with distinct demands e.g., security, latency, resiliency, and bandwidth. To deal with this challenge, the NGMN has proposed the concept of "5G network slicing". It supports the main architecture framework for 5G which permits for various 5G network slicing architectures that also involves shared network, cloud, and virtualized functions resources to exist in parallel. A multi-tenant 5G network looks forward to a slicing architecture that dynamically provides virtual 5G network slices addressing requirement and virtualized SDN/NFV control instances for a full tenant control of allocated virtual resources [10].

Full multi-tenancy-based network sharing expansion depends on capacities of software-based and virtualization mechanisms those are gradually presented into 3GPP network, prompting its standardization roadmap.

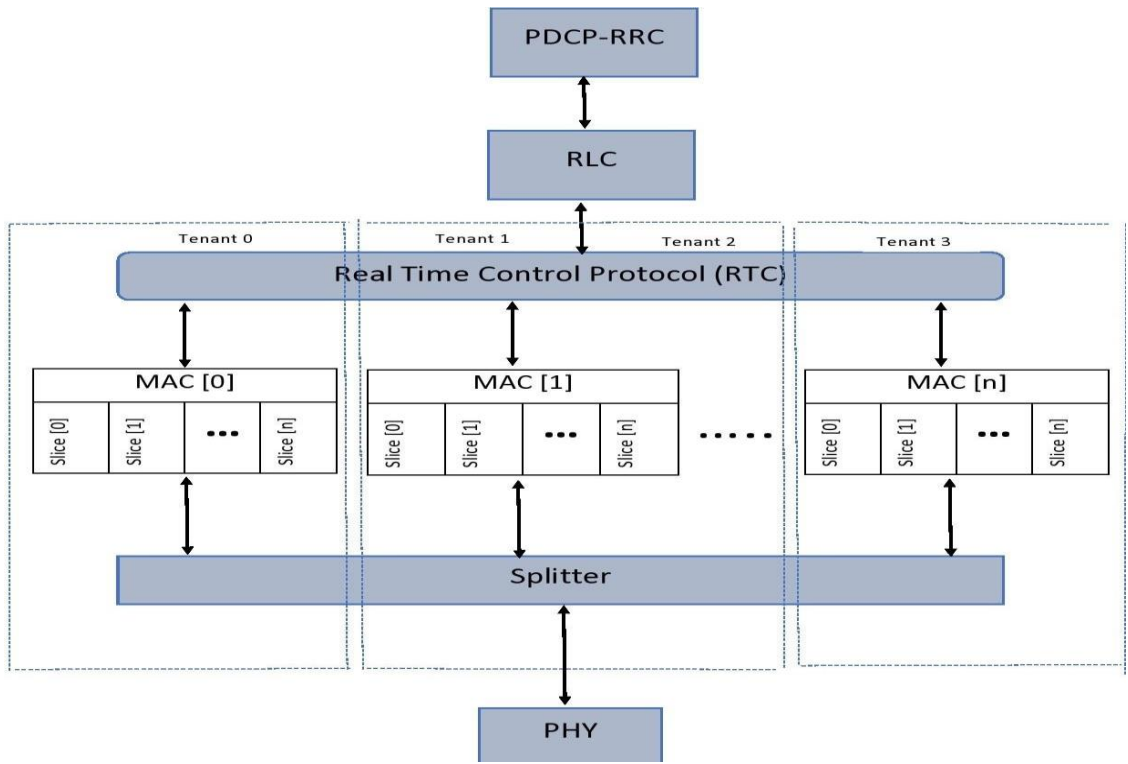

Figure 5. Architecture of multi-level slicing

The above figure is the whole architecture of multi-level slicing. The dotted boxes are independent operators residing in the same base station. Inside each of them traffic can further be divided or sliced based on the application e.g., VOIP, video etc. In this scenario, we call the operators as tenants. On top of the tenants, there is one real time controller that controls the tenants or operators and all the spectrum sharing techniques work here. When packets arrive from RLC layer, the destination UE id is extracted and search for the corresponding MAC id. The packets are then sent to the appropriate MAC and to the appropriate sub-slice according to their corresponding slice id.

The PDCP-RRC module links up NIC and LTE IP modules. This module collects data in downstream direction from upper layer and in upstream manner from the RLC layer whereas the RLC module executes de-multiplexing and multiplexing pf MAC SDUs to/from the MAC layer. RLC operation is the same on both eNB and the UE. Again, the MAC module takes care of buffering and processing the MAC SDUs and MAC PDUs coming from other layers along with scheduling, resource allocation and Adaptative Modulation and Coding (AMC). On each TTI specific links are scheduled for transmission following the schedule list provided by the scheduler. The PHY module implements functions related to the physical layer such as channel feedback computation and reporting, data transmission and reception and control messages handling. This multi-level slicing has done on modified SimuLTE model using Omnet++.

Currently research has been going on to combine IoT with 5G network technology. Machine learning algorithm based automated network slicing has been suggested to accommodate network requirement according to the vendor usage and different scenarios of the organization. Slicing concept of 5G network is considered as an efficient way to enhance the security of IoT devices [12]. This research work has not covered the security issues of 5G networks. RecentlyBlockchain is using brokering mechanism to ensure private and secure transactions between the resource provider for 5G services and the network slice provider [13,14].

## 5. DISCUSSION

In single level slicing, allocation is static which means different types of services (audio, video, text etc.) cannot be provided in same MAC. Different MAC is used to provide different type of services. On the other hand, multilevel slicing supports dynamic allocation which means a single MAC can be used to provide different types of services in parallel. Along with this feature different types of services can use different amount of resource blocks according to their requirement. For example, if there are 25 recourse blocks; audio service is using 10 resource blocks and video is using 15 recourse blocks.

## 6. CONCLUSION

Finally, this paper presented active RAN with multi-level slicing with a new system level simulation model that can simulate multi-tenant RANs. The advantage of fine-grained based spectrum sharing is highlighted by analyzing the prior results along with the simulation platform presentation in details.

Subsequently, we are planning to work on the spectrum sharing technique includes both time sharing algorithm and frequency sharing algorithm where we want to analyze their impact on spectrum sharing granularities. Correspondingly we will compare the multi-level sharing with different sharing technique based on multi-level sharing to observe some interesting outcome.

ACKNOWLEDGEMENTS

The authors would like to thank everyone, just everyone!

## AUTHORS

**Wardah Saleh** received the B.Sc degree in computer science and engineering from American International University-Bangladesh, Dhaka, Bangladesh, in 2017 and the M.Sc. degree in computer networks and architecture from American International University Bangladesh, Dhaka, Bangladesh, in 2018. She is currently pursuing her carrier as a Lecturer at the department of Computer Science in America International University Bangladesh, Dhaka, Bangladesh from 2018.

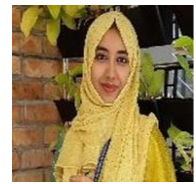

**Shahrin Chowhdury** received the B.Sc. degree in computer science from American International University-Bangladesh, Dhaka, Bangladesh, in 2006 and the M.Sc. degree in networks and distributed systems from Chalmers University of Technology, Gothenburg, Sweden, in 2013. She is currently pursuing her carrier as an Assistant Professor of Computer Science department of American International University-Bangladesh.

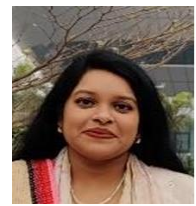